\documentstyle[11pt]{article}
\begin{document}
\title{Running Surface Couplings}
\author{Sergei D. Odintsov\\
Tomsk Pedagogical Institute, 634041 Tomsk, Russia and \\
Facultad de Fisica, Universidad de Barcelona, \\
Diagonal 647, E-08028, Barcelona, Spain\\
\\
A. Wipf\\
Theoretisch-Physikalisches-Institut, FS-Universit\"at Jena,\\
Max-Wien-Platz 1, 07743 Jena, Germany}
\date{20.5.1995\\
preprint uni-jena-TPI 1/95}
\maketitle

\newcommand{\eqnn}[1]{\begin{eqnarray*}#1\end{eqnarray*}}

\newcommand{\eqnl}[2]{\par\parbox{11cm}
{\begin{eqnarray*}#1\end{eqnarray*}}\hfill
\parbox{1cm}{\begin{eqnarray}\label{#2}\end{eqnarray}}}

\newcommand{\eqnlb}[2]{\begin{equation}\fbox{$\displaystyle#1
$}\label{#2}\end{equation}}

\newcommand{\eqngr}[2]{\par\parbox{11cm}
{\begin{eqnarray*}#1\\#2\end{eqnarray*}}\hfill
\parbox{1cm}{\begin{eqnarray}\end{eqnarray}}}

\newcommand{\eqngrlb}[3]{\par\parbox{11cm}
{\begin{eqnarray}\fbox{$\displaystyle#1\\#2$}\end{eqnarray}}\hfill
\parbox{1cm}{\begin{eqnarray}\label{#3}\end{\eqnarray}}}

\newcommand{\eqngrl}[3]{\par\parbox{11cm}
{\begin{eqnarray*}#1\\#2\end{eqnarray*}}\hfill
\parbox{1cm}{\begin{eqnarray}\label{#3}\end{eqnarray}}}

\newcommand{\eqngrr}[3]{\par\parbox{11cm}
{\begin{eqnarray*}#1\\#2\\#3\end{eqnarray*}}\hfill
\parbox{1cm}{\begin{eqnarray}\end{eqnarray}}}

\newcommand{\eqngrrl}[4]{\par\parbox{11cm}
{\begin{eqnarray*}#1\\#2\\#3\end{eqnarray*}}\hfill
\parbox{1cm}{\begin{eqnarray}\label{#4}\end{eqnarray}}}

\newcommand{\eqngrrrl}[5]{\par\parbox{11cm}
{\begin{eqnarray*}#1\\#2\\#3\\#4\end{eqnarray*}}\hfill
\parbox{1cm}{\begin{eqnarray}\label{#5}\end{eqnarray}}}

\newcommand{\eqngrrrrl}[6]{\par\parbox{11cm}
{\begin{eqnarray*}#1\\#2\\#3\\#4\\#5\end{eqnarray*}}\hfill
\parbox{1cm}{\begin{eqnarray}\label{#6}\end{eqnarray}}}

\newcommand{\refs}[1]{(\ref{#1})}

\def\mtxt#1{\quad\hbox{{#1}}\quad}
\def\eps{\epsilon}
\def\ov{\over}
\def\ha{{1\over 2}}
\def\xis{\xi\!-\!{1\ov 6}}
\def\lam{\lambda}

\abstract{
We discuss the renormalization group improved
effective action and running surface couplings in curved
spacetime with boundary. Using scalar self-interacting
theory as an example, we study the influence of boundary
effects to effective equations of motion in spherical
cap and the relevance of surface running couplings to quantum cosmology
and symmetry breaking phenomenon.
Running surface couplings in the asymptotically free SU(2)
gauge theory are found.
}

\vfill\eject

\section{Introduction}
\vskip .5truecm
Boundary terms may play an important role in quantum cosmology
and in particular in connection with the quantum state of the
universe \cite{HH}. That is why,
starting from the 80'ties \cite{KCD}, there has been a continued
interest to study boundary divergences (see, for
example, [2-6,8] and references therein). In \cite{V,MP} some
misprints of previous calculations have been corrected and the
surface divergences have been found in a form of conformal anomalies for
various boundary conditions.

In \cite{O} the running surface couplings have been introduced. The
motivation to do it was the fact that in order to make a theory
multiplicatively renormalizable in curved spacetime with boundary
one has to include the surface Lagrangian with arbitrary coupling constants
in the total Lagrangian. When
the renormalization group is constructed, each coupling becomes a running
effective coupling. A similiar idea has been persued in \cite{WW},
where running surface couplings have been discussed in spacetime with
boundaries and have been related to the finite size effects.
It is quite well-known that running couplings have
different physical applications. It is the purpose of this work to discuss
the running surface couplings for different theories and to look
for the consequences to which they may lead.

In the next section we discuss the self-interacting scalar theory on curved
spacetime with boundary using Dirichlet boundary conditions. The explicit
expressions for the volume and running surface couplings are given. The
procedure to construct the RG improved effective action in such a
spacetime is discussed. In the section 3 we find the
RG improved effective action in a spherical
cap and show how boundary terms become relevant in the effective field
equations. For the example of a disc we show the
possible influence of boundary terms to symmetry breaking phenomena. In
section 4, we show how the above discussion can be generalized to arbitrary
GUTs, and in particular to the asymptotically
free SU(2) gauge theory with scalars and spinors,
in curved spacetime with boundary. Some discussions
are presented in the last section.
\section{Self-interacting scalar theory in curved space with boundary.}
Consider the self-interacting scalar theory in curved spacetime
$M$ with boundary $\partial$M.
The renormalization of the theory maybe done in close analogy with
the renormalization in curved spacetime without boundary
(for a general introduction see \cite{BOS}). The boundary
conditions for scalar fields maybe chosen to be of Dirichlet type
\eqnl{\phi(x)=0 \mtxt{,} x \in \partial M}{gl1}
or Robin type
\eqnl{(\psi+n^{\mu}\nabla_{\mu})\,\phi(x)=0 \mtxt{,}x \in \partial M.}{gl2}
Here $n^{\mu}$ is the outward normal on $\partial M$ and $\psi$ is an arbitrary
scalar function.

The euclidean action corresponding to a massless multiplicatively
renormalizable theory maybe written as the following:
\eqnl{S=S_M+S_V+S_S,}{gl3}
where
\eqngrl{S_M&=&\int d^4x\sqrt{g}\Big\{{\ha g^{\mu
\nu}\partial_{\mu}\varphi\partial_{\nu}\varphi+\ha\xi
R\varphi^2+\frac{{\lam \varphi}^4}{4!}}\Big\}\;,}
{S_V&=&\int d^4x\sqrt{g}\Big\{{a_1R^2+a_2C^2_{\mu
\nu\alpha\beta}+a_3G+a_4\Box R}\Big\},}{gl4}
and $C_{\mu\nu\alpha\beta}$ is the Weyl tensor, G the Gauss-Bonnet
invariant and $a_1,a_2, a_3, a_4$ are coupling constants in
the external fields sector.

In the discussion of the surface action we will limit ourselves to Dirichlet
boundary conditions. We use two invariants of dimension $L^{-3}$
expressed in terms of $R_{\mu\nu\alpha\beta}$ and the extrinsic
curvature of the boundary $K_{\mu\nu}$ \cite{BG,MP}
\eqngrrl{q&=&\frac{8}{3}K^3+\frac{16}{3} K_{\mu}^{\;\nu}
K_{\nu}^{\;\alpha}K_{\alpha}^{\;\mu}-8 K K_{\mu\nu}K^{\mu\nu}+4KR}
{&-&8R_{\mu\nu}(Kn^{\mu}n^{\nu}+K^{\mu\nu}) +
8R_{\mu\nu\alpha\beta} K^{\mu\alpha}n^{\nu} n^{\beta},}
{g&=&K_{\mu}^{\;\nu}K_{\nu}^{\;\alpha}K_{\alpha}^{\;\mu}-K
K_{\mu\nu}K^{\mu\nu}+\frac{2}{9}K^3.}{gl5}
Then, the surface action maybe rewritten as
\eqnn{S_S=\int\limits_{\partial M}d^3x\,\sqrt{\gamma}\,L_S \;}
with
\eqnl{
L_S=\alpha_Dq+\beta_Dg+\gamma_DRK+\delta_D n^\mu
\nabla_\mu R+\zeta_D C_{\mu\nu\alpha\beta}K^{\mu\alpha}n^\nu n^\beta,}{gl6}
where $\gamma_{\alpha\beta}$ is the induced metric of the
boundary and $\alpha_D,\dots,\zeta_D$ are surface coupling constants. In the
same way one can write $S_S$ for other boundary conditions.

Now, from the point of view of the renormalization group, each coupling
constant has the correspondent effective coupling constant.
Using the well-known
results for the one-loop divergences of the volume terms one easily finds the
running volume couplings:
\eqngrrl{
\lam (t)&=&{\lam  \ov \kappa(t)}\qquad\qquad ,\qquad\qquad
\xi(t)={1\ov 6}+\big(\xis\big)\kappa(t)^{-{1\ov 3}}}
{a_1(t)&=&a_1-{1\ov 2\lam }\big(\xis\big)^2
\big[\kappa(t)^{1\ov 3}-1\big]\quad,\quad
a_2(t)=a_2+{t\ov 120(4\pi)^2}}
{a_3(t)&=&a_3-{t \ov 360(4\pi)^2}\:\:,\:\:
a_4(t)=a_4-{t\ov 180(4\pi)^2}-{\xis\ov 12\lam }\big[
\kappa(t)^{2\ov 3}-1\big],}{gl7}
where $t$ is renormalization group parameter and
\eqnn{
\kappa(t)= 1-{3\lam t\ov (4\pi)^2}.}
Using the explicit results
for the boundary conterterms \cite{BG,MP} we can write down the explicit
expressions for the running surface couplings in theory
\refs{gl3} \cite{O,WW}:
\eqngrl{
\alpha_D(t)&=&\alpha_D-{t\ov 360(4\pi)^2}\qquad,\qquad
\beta_D(t)=\beta_D+{2t\ov 35(4\pi)^2}}
{\gamma_D(t)&=&\gamma_D+{D(t)\ov 3}\;,\,
\delta_D(t)=\delta_D+{D(t)\ov 2}\,,\,
\zeta_D=\zeta_D+{t\ov 15(4\pi)^2}}{gl8}
where
\eqnn{
D(t)={\xis \ov 2\lam }\big[\kappa(t)^{2/3}-1\big].}
As usually the $t\to\infty$ limit defines the theory at very high energies
(strong
gravitational field). As we see from Eqs. \refs{gl8} there is already some
mixture of the volume with the surface couplings when they are running.

Now, after this overview of the situation with running surface couplings in
curved spacetime, the interesting question is -- what new phenomena may
be encountered using
the renormalization group. In particular, as it was already mentioned, the
boundary effects are expected to be important in quantum cosmology. Hence it
is interesting to understand the relevance of renormalization group  in this
respect.

Let us consider the situation where the volume Lagrangian (as well as $L_S$)
is independent of one of the coordinates. Then, in the volume
action we may integrate explicitly over this coordinate and as a result we
can write the action (assuming that there is only a gravitational
background field) as
\eqnl{
S_{grav.}=\int d^3\,\sqrt{g}\,\Big\{ l_1L_V+l_2L_S\Big\},}{gl9}
where $l_1, l_2$ are some dimensionful constants, for example,
$l_1=\int dx$ (where $x$ is the variable on which the Lagrangean
does not depend). Due to the fact that the theory is multiplicatively
renormalizable, we may now write explicitly the RG equation for effective
Lagrangian:
\eqnl{
(\mu{\partial\ov \partial\mu}+\beta_i{\partial\ov\partial\lam _i}
-\gamma_i \phi_i{\delta\ov\delta\phi_i})\;L_{\mbox{eff}}\;(\mu,\lam _i
\,,\,\phi_i)=0,}{gl10}
where $\mu$ is a mass parameter, $\lam _i$ are volume and
surface coupling constants
with corresponding beta-functions $\beta_i$ and $\phi_i$
are the fields. For an alternative derivation of \refs{gl10},
where $\mu$ is replaced by the inverse diameter of the
spacetime $M$ see \cite{WW}.

Solving Eq.\refs{gl10} by the method of characteristics, with
Lagrangean \refs{gl9} as initial condition at $t=0$ and
assuming a gravitational background field only (the other
background fields are set to zero) we find the
following contribution to $L_{eff}$
\eqngrrrl{&&L_{eff}(\mu,\lam _i,\phi_i)=
L_{eff}(\mu e^t,\lam _i(t),\phi_i(t))}
{&&\qquad = l_1\Big\{ a_1(t)R^2+a_2(t)C^2_{\mu\nu\alpha\beta}
+a_3(t)G+a_4(t)\Box R\Big\}}
{&&\qquad+l_2\Big\{\alpha_D(t)q+\beta_D(t)g+\gamma_D(t)RK+
\delta_D(t)n^{\mu}\nabla_{\mu}R}
{&&\qquad+\zeta_D(t)C_{\mu\nu\alpha\beta}K^{\mu\alpha}n^{\nu}n^{\beta}\Big\},}
{gl11}
where the running volume and surface couplings are given by
eqs.(\ref{gl7},\ref{gl8}). The above discussion which yielded the RG improved
Lagrangian in curved space is very similar to standard RG
improvement of the
effective potential in flat \cite{CW,S} or in curved space
\cite{BVW,EO}. The problem
now is the choice of RG parameter $t$. Motivated by the one-loop
considerations of the theory under discussion, the natural choice is
(let R be positive)
\eqnl{t=\ha\log\frac{R}{\mu^2}.}{gl12}
With this choice, we get the improved effective Lagrangian
(the summation over all leading logarithms of perturbation theory).
In that sense the result is beyond one-loop order. The important
implication of (\ref{gl11},\ref{gl12}) is that due to the RG, the surface
terms cease to be surface terms. They give contributions
to the equations of motion, and hence, they influence quantum cosmology
dynamically. Classically the surface terms maybe dropped. On the
quantum level, however, these terms are important, as after RG improvement they
contribute
         to the equations of motion. We give an explicit example in the next
section.

\section{RG improved Lagrangian in scalar theory on four-sphere with
boundary}
In what follows we will limit ourselves to the spaces of the type
$R_{\mu\nu}=\Lambda g_{\mu\nu}$ which are of interest for quantum
cosmology as they describe the inflationary Universes. In this case the
structure of the initial Lagrangian significally simplifies.

Consider as an example a spherical cap $C$, i.e. region of the four-sphere
with maximum colatitude $\theta$. Then, the RG improved action is
\eqngrrl{
S_{eff}&=&\int\limits_M d^4x\sqrt{g}\,S_{V,eff}+\int\limits_{\partial
M}d^3x\sqrt{\gamma}\,S_{S,eff}}
{&=&24\pi^2\Bigg\{\Big[16a_1(t)+{8\ov 3}a_3(t)\Big]\cdot
\Big[{1\ov 2}-{3\ov 4}\cos\theta+{1\ov 4}\cos^3\theta\Big]}
{&+&\cos^3\theta\Big[{2\alpha_D(t)}\Big]
+{9\ov 2}\cos\theta\sin^2\theta\big\{ \gamma_D(t)\big\}\Bigg\},}{gl13}
where $t={1\ov 2}\log{4\Lambda\ov\mu^2}$. We supposed Dirichlet boundary
conditions for the scalar field. One may consider other conditions as
well. The calculation   of conformal anomaly in above-described
situationes has
been given in \cite{MP}. For comparison we may give the RG improved action in
case of the 4-sphere(for the discussion of the effective
action in De Sitter space see, also \cite{FT} and \cite{CDGR})
\eqnl{S_{eff}=24\pi^2\Big(16a_1(t)+{8\ov 3}a_3(t)\Big).}{gl14}
The effective equations of motion are given by
\eqnl{\frac{\partial S_{eff}}{\partial\Lambda}=0.}{gl15}
Classically $a_1$ and $a_3$ are constant and the cosmological constant
is not determined. On quantum level we get from from
(\ref{gl14},\ref{gl15})
\eqnn{
8\Big(\xis\Big)^2\kappa(t)^{-2/3}-{1\ov 135}=0,}
where $\kappa(t)$ has been introduced below \refs{gl7}, the
selfconsistent quantum solution
\eqnn{
{1\ov 2}\log{4\Lambda\ov \mu^2}={(4\pi)^2\ov 3\lam}\Bigg\{
1-\Big[8\cdot 135(\xis)^2\Big]^{3/2}\Bigg\}.}
Hence, the effective cosmological constant is defined from the
back-reaction of the quantum matter on the geometry. The corresponding
non-singular universe is a De-Sitter spacetime (for free theory
see also\cite{DC}).

Let us now consider a universe which is a spherical
cap $C$. Its RG improved gravitational action is given by
\refs{gl13}. The effective equation is found to be
\eqngrl{
&&\Big[8(\xis)^2\kappa(t)^{-2/3}-{1\ov 135}\Big]
\Big[{1\ov 2}-{3\ov 4}\cos\theta+{1\ov 4}\cos^3\theta\Big]}
{&&-{2\cos^3\theta\ov 360}-{3\ov
2}\cos\theta\sin^2\theta(\xis )\kappa(t)^{-1/3}=0.}{gl16}
This effective equation of motion in which the boundary
effects have been taken into account, cannot be solved explicitly.
Assuming $\lam t$ (on which $\kappa$ depends) to be small
and keeping only terms which are linear
in this parameter we get the quantum solution
\eqngrrl{
&&-{1\ov 2}\log{4\Lambda\ov\mu^2}=
\Bigg\{\Big[8(\xis)^2-{1\ov 135}\Big]\Big[\ha
-{3\cos\theta\ov 4}
+{\cos^3\theta\ov 4}\Big]}{&&\qquad\qquad
-{\cos^3\theta\ov 180}-{3\ov 2}\cos\theta\sin^2\theta(\xis)\Bigg\}}
{&&\cdot\Bigg\{{16(\xis)^2\lam\ov (4\pi)^2}\Big[
\ha-{3\cos\theta\ov 4}+{\cos^3\theta\ov 4}\Big]-{3\ov 2}\cos\theta
\sin^2\theta(\xis){\lam\ov (4\pi)^2}\Bigg\}^{-1}.}{gl17}
As one sees the boundary terms play an important role. They change
the structure of the self-consistent effective equation
qualitatively. Our considerations provides an example how
through the RG the boundary terms may become relevant in
quantum cosmology.

Moreover, this feature is quite general and maybe extended to
any renormalizable theory - this only changes the coefficients
in \refs{gl13} and possibly $\gamma(t)$. One may further
admit a scalar background field in which case $L_{eff}$ becomes
quite complicated and leads to two sets of effective equations of motion.

As another application one can consider the wave function of the Universe
\cite{HH} which is defined (in our example) as path integral with a
spherical cap as boundary surface
\eqnl{\psi(\Lambda)=e^{-S_{eff}}.}{gl18}
The solution of the field equations is given by \refs{gl17} and yields
the curvature $R=4\Lambda$ of such a spacetime or equivalently
its radius $R=\frac{1}{a^2}$. The effective action is the obtained
by substituting \refs{gl17} into \refs{gl13} and with \refs{gl18}
yields to the wave function of the system and to
       the probability distribution on the set of boundary conditions.

As an another interesting example let us consider a ball $D$, i.e. the region
in flat spacetime bounded by a three-sphere. We suppose that the
scalar background is non-zero and constant. Then we may calculate
$S_{eff}$ in \refs{gl11} as the follows:
\eqnl{
S_{eff}=V_4\cdot\Big\{{\lam (t)\varphi^4\ov 4!}-2c_1\alpha_D(t)\Big\}}{}

where $V_4$ is 'volume' of the ball and $c_1$ is a
dimensionless constant. It is evident that in this case
$t=\ha\log\varphi^2/\mu^2$, as in Coleman-Weinberg approach
\cite{CW}. Now one may discuss the symmetry breaking induced by boundary
effects (for the first study of symmetry breaking under
external curvature ,see \cite{SHO}).
         Solving the equation of motion
$\frac{\delta S}{\delta\varphi}=0$ to first order in $\lam $ we get
\eqnl{\varphi^4={c_1 \ov 120\lam (4\pi)^2}.}{gl21}
Classically $\varphi=0$, and no symmetry breaking occurs. This simple example
shows how boundary effects may trigger the spontaneous symmetry
breaking. Now we turn to the discussion of more complicated theories.

\section{Running surface constants in GUTs.}
Let us show now that one can easily generalize the above picture to
the (for simplicity) massless GUT's in curved spacetime. We will
consider an arbitrary asymptotically free GUT (for a list of such GUTs, see
for example \cite{VT}). In this case, we have for running gauge,
Yukawa and scalar couplings
\eqnl{g^2(t)=\frac{g^2}{1+a^2g^2t}\quad,\quad h^2(t)=k_2g^2(t)\mtxt{and}
f(t)=k_1g^2(t),}{gl22}
where for           Yukawa and scalar couplings $k_1$ and $k_2$
are constant matrices. The scalar-gravitational running coupling
is generally of the form \cite{BOS}
\eqnl{
\xi(t)={1\ov 6}+(\xis)(1+a^2g^2t)^B,}{gl23}
where $B$  maybe positive or negative, depending on the
detailed field-content of the theory. The running volume
couplings have the structure similar as those
in section 1 (powers of terms connected with $\xi$ are changing
according to \refs{gl23}), so we will not present them here (for details, see
\cite{BOS}). As regards to the running surface couplings they maybe
easily found using the general results of refs. \cite{BG,MP}.
To be more specific let us consider the asymptotically free SU(2)
gauge theory with one scalar and two spinor triplets \cite{VT}.
Imposing the boundary conditions of refs. \cite{GJ,WW} for the
fermions and absolute boundary conditions for the scalars and gauge fields
and assuming $R_{ab}=\Lambda g_{ab}$ we find now the boundary action
\eqngrrl{
S_S&=&\int\limits_{\partial M}d^3 x\,\sqrt{\gamma}\,L_S}
{L_S&=&\beta_1\Lambda
K+\beta_2K^3+\beta_3KK_{\mu\nu} K^{\mu\nu}}
{&&\quad+\beta_4 K_\mu^{\;\nu}K_\nu^{\;\alpha}K_\alpha^{\;\mu}+\beta_5
C_{\mu\nu\alpha\beta}K^{\mu\alpha}n^{\nu}n^{\beta},}{gl24}
where the corresponding running couplings are
\eqngrrrrl{
\beta_1(t)&=&\beta_1-{t\ov (4\pi)^2}({62n_A\ov 135}-{11n_F\ov 135})-
{4(\xis)\ov 3(4\pi)^2(B +1)a^2}[(1+a^2t)^{B+1}-1]}
{\beta_2(t)&=&\beta_2+{t\ov (4\pi)^2}({n_s\ov 27}+{17n_F\ov 945}
-{338n_A \ov 945})}
{\beta_3(t)&=&\beta_3+{t\ov (4\pi)^2}({n_s\ov 45}+{13 n_F\ov 315}+
{58n_A\ov 63})}
{\beta_4(t)&=&\beta_4+{t\ov (4\pi)^2}({4n_s\ov 135}-{116n_F\ov 945}-
{436n_A\ov 945})}
{\beta_5(t)&=&\beta_5+{t\ov (4\pi)^2}({2n_s\ov 45}-{7n_F\ov 45}-{26n_A\ov
 45}),}{gl25}
where for the SU(2) model $n_A=3, n_s=3, n_F=3$ or $n_F=6$ and \cite{BO}
\eqnn{
\xi(t)={1\ov 6}+\big(\xis)(1+a^2g^2t\big)^
{ -\big( { 12-{5\ov 3}k_1-8k_2  \ov b^2 } \big) }.}
Here $b^2$ is constant and $k_1,k_2$ can be found in \cite{VT}. For
$n_F=3$ we have
\eqnn{
\Big(12-{5\ov 3}k_1-8{k_2\ov b^2}\Big)<0}
and for $n_F=6$ we have $B>0$. The running
surface couplings in other GUTs can be found similarly
as for the scalar theory considered in the previous section. They
lead to corrections of the quantum states in quantum cosmology.
\section{Conclusion.}
We have discussed RG improved effective action in curved
spacetime with boundaries. The running surface couplings
are getting important in this approach as they maybe
relevant in different physical applications. Among examples
given in this work we have studied the influence
of the boundary terms to the effective field equations,
possible application to quantum cosmology and symmetry
breaking. Note that we have studied all these questions
using the effective action on constant curvature spaces.
Nevertheless, one may apply similar technique to the non-local
effective action and black hole physics where boundary
effects may also play an important role. We hope to return
to some of these questions in near future.

\vfill\eject

\end{document}